# Population aging caused by rise in sex ratio at birth

Zhen Zhang[1]

Qiang Li[2]

**Abstract**

Despite its historical and biological stability, the sex ratio at birth (SRB) has risen in parts of world in the last several decades. The resultant demographic consequences, mostly on sex imbalance, are well documented, typically including "missing girls/women" and "marriage squeeze." However, the SRB-induced impact on demographic dynamics, particularly its underlying mechanism, has not been explored in depth. We aim to investigate the impact of the SRB rise on the size, structure, and growth of a population, particularly emphasizing on population aging. We provide a simple framework, derived from classical stable population models, to analyze how the SRB rise can reduce the population size and make the population old. We demonstrate that the cohorts born with a higher SRB are smaller in size than those with a lower SRB. As the affected cohorts are born into the population, their smaller size will reduce the total population size, thereby lifting the fraction of old people that were born with the original SRB and have the same size as before. The resultant population aging speed increases as the cohorts with the new SRB take an increasing share of the population. This study adds that, in addition to fertility and mortality, the SRB can be a driving factor of population dynamics, especially when it moves far above normal biological levels.

---

[1] Institute of Population Research, School of Social Development and Public Policy, Fudan University, Shanghai, China.

[2] Population Research Institute, School of Social Development, East China Normal University, Shanghai, China. Email: qli@soci.ecnu.edu.cn

# 1. Introduction

The human sex rate at birth (SRB) is biologically stable at around 1.05 male births per female birth (Chahnazarian 1988). Since the late 1970s, however, SRBs have risen in a number of countries in Asia and eastern Europe (Babiarz et al. 2019; Chao et al. 2019; Guilmoto 2012a; Guilmoto and Ren 2011; Cai and Lavely 2003), due to prenatal sex selection driven by rapid fertility decline, strong son preferences, and easy access to sex selection technologies (Guilmoto 2009, 2012a, 2012b, 2012c; Kashyap and Villavicencio 2016; Meslé et al. 2007). As concerns are mounting on the rapid rise in the SRB and the relevant socio-demographic consequences, efforts have been made to lower the abnormally high SRB, including introducing or strengthening national policies and programmes in addressing gender inequality and sex selection at birth (Guilmoto 2012a; Li 2007). In recent years, South Korea has reduced its SRB to biologically normal levels (Den Boer and Hudson 2017), and some other countries such as China has seen the slowing down of the SRB rise (Chao et al. 2019).

The resultant demographic consequences, typically including "missing girls/women" and "marriage squeeze," are well documented (Branches 2004; Guilmoto 2012a, 2015; Jiang et al. 2014; Li 2007; Sen 1990; Tuljapurkar et al. 1995). Besides, several studies, based on simulation results, reported that a higher SRB often comes with population aging or slowing population growth (Cai and Lavely 2003; Chen and Zhang 2019; Chen and Li 2010; Jiang et al. 2011), but few have explored the underlying mechanism in depth. It thus remains unknown, whether the link between sex ratios and population aging is coincident under specific population conditions, and if not, how it occurs and what factors are involved in the process.

This paper aims to investigate the impact on the population structure of the increase in birth masculinity. In doing so, we derive a simple framework from classical stable population models (Lotka 1939). Within this framework, we apply a perturbation analysis to examine how a rise in the SRB reduces the population size and causes population aging.

The rest of this paper is organized as follows. First, we introduce the settings and notation used in this study. Second, we begin with a simple case in which the number of births is constant over time. We formulate the relationships between the SRB and population size and visualize the process of population aging resultant from the SRB rise. Third, we present the case with constant fertility schedules, which is not only more realistic, but also can show the pure impact of the rise in the SRB. Finally, we discuss the implications of our findings.

# 2. Settings and notation

We base our analysis on the assumption of female dominance; that is, that population dynamics are determined by female vital rates and there are always sufficient males to match all of the females. This assumption can be justified by the widespread excess males in all populations with the abnormally high SRB (Guilmoto 2012a, 2012b, 2015; Sen 1990; Tuljapurkar et al. 1995). Additionally, the population is assumed to be closed to migration. For simplicity, we use the following notation:

- $SRB(t)$: the sex ratio at birth at time $t$, expressed by the number of male births per female birth.
- $\pi_f(t), \pi_m(t)$: the proportions of female and male births at time $t$, with $\pi_f(t) +$

$\pi_m(t)=1$. The two proportions are equivalent to SRB; for example, $\pi_m(t) = SRB(t)/(1 + SRB(t))$ or $SRB(t) = \pi_m(t)/(1 - \pi_m(t))$.

- $N_f(a,t), N_m(a,t), N(a,t)$: female, male, and total populations aged $a$ at time $t$.
- $B_f(t), B_m(t), B(t)$: female, male, and total births at time $t$. The number of female and male births is given by $B_s(t) = \pi_s B(t), s = f, m$, respectively. If annual births are fixed, then $B(t) = B$.
- $\omega$: upper limit of human lifespan.
- $p_f(a), p_m(a)$: constant probability that a newborn girl (boy) will survive to age $a$.
- $p(a)$: constant probability that a newborn child will survive to age $a$, which is the weighted average of $p_f(a)$ and $p_m(a)$: $p(a) = \pi_m \cdot p_m(a) + (1 - \pi_m) \cdot p_f(a)$.
- $e_f^o, e_m^o, e^o$: life expectancies at birth for females, males, and the total population, defined by the integral of $p(a)$ from age 0 to $\omega$; for instance, $e^o = \int_0^\omega p(a)da$. The three life expectancies are related to one another via $e^o = \pi_m \cdot e_m^o + (1 - \pi_m) \cdot e_f^o$, usually with $e_m^o < e_f^o$.
- $\delta(a) = p_f(a) - p_m(a)$: the sex difference in the probability of surviving from birth to age $a$, which is usually positive due to female survival advantages. We then have $\int_0^\omega \delta(a)\, da = e_f^o - e_m^o$, which is the sex gap in life expectancy at birth.
- $m(a)$: constant fertility schedules of women for female and male offspring combined, with the total fertility rates $TFR = \int_\alpha^\beta m(a)da$, where $\alpha$ and $\beta$ are the lower and upper bounds of childbearing age, respectively.
- $m_f(a), m_m(a)$: constant fertility schedules for female and male offspring, respectively.

## 3. SRB-induced reduction in population size

In this section, we demonstrate how the rise in the SRB can reduce population size. Consider a female-dominant stationary population that is characterized with a fixed number of annual births, a fixed life table for each sex, and zero net migration rates at all ages. Hence, the number of persons aged $a$ at time $t$ equals the number of births $t - a$ years earlier times the probability of surviving from birth to age $a$:

$$N(a,t) = B(t-a) \cdot p(a) = B \cdot p(a). \tag{1}$$

The total population size is thus given by

$$N(t) = \int_0^\omega N(a,t)da = B \cdot \int_0^\omega p(a)da = B \cdot e^o \tag{2}$$

where $e^o$ is life expectancy at birth for the population as a whole.

For a two-sex population, $N(a,t)$ can be expressed in terms of the proportion of male births $\pi_m$:

$$N(a,t) = B \cdot \left(\pi_m \cdot p_m(a) + (1 - \pi_m) \cdot p_f(a)\right)$$

$$= \pi_m \cdot B \cdot \left(p_m(a) - p_f(a)\right) + B \cdot p_f(a)$$

$$= -\pi_m \cdot B \cdot \delta(a) + B \cdot p_f(a). \tag{3}$$

The derivative of Eq. (3) with respect to $\pi_m$ is

$$\frac{dN(a,t)}{d\pi_m} = -B \cdot \delta(a). \tag{4}$$

The slope at age $a$, with the sign reversed, is the product of $B$ and $\delta(a)$. This expression indicates how many more of the extra male births resulting from a rise in $\pi_m$ would be removed up to age $a$ because of male survival disadvantage.

Analogously, the total population can be expressed as:

$$N(t) = B \cdot e^o = B \cdot \left(\pi_m \cdot e_m^o + \pi_f \cdot e_f^o\right) = -B \cdot \pi_m \cdot \left(e_f^o - e_m^o\right) + e_f^o. \tag{5}$$

Differentiating Eq. (5) gives

$$\frac{dN(t)}{d\pi_m} = -B \cdot \left(e_f^o - e_m^o\right). \tag{6}$$

Since $e_f^o > e_m^o$, the equation is negative, indicating the total number of extra deaths that would occur if more boys were born due to the rise in $\pi_m$.

The mechanism underlying such impact is intuitively straightforward. Given the number of births, the higher the SRB, the more boys and the fewer girls are born. Since males are usually subject to higher mortality than their female counterparts, more boys imply more deaths and thus lead to the reduction of population size. As shown in Eq. (5), with births $B$ fixed and $\pi_m$ rising, increasing weight of male population will lead to decreasing $e^o$ and, hence, population.

The impact of the SRB is symmetrical: a decrease in $\pi_m$ or equivalently a rise in $\pi_f$ can enlarge population size, unless the resulting changes will invalidate the assumption of female dominance, and vice versa. Indeed, the following results derived in this study about the impact of the change in the SRB on population size and structure can work in both directions, as in Eqs. (4) and (6).

Dividing Eq. (4) by Eq. (1) gives the rate of the relative change of $N(a,t)$ given a rise in $\pi_m$:

$$\frac{d \ln N(a,t)}{d\pi_m} = \frac{dN(a,t)}{N(a,t)d\pi_m} = -\frac{B \cdot \delta(a)}{B \cdot p(a)} = -\frac{\delta(a)}{p(a)}. \tag{7}$$

Equation (7) shows that, with the removal of the impact of the number of births, the resultant reduction in population size depends on the ratio of sex difference in the probability of surviving from birth to age $a$ to the same probability but for the population as a whole. In the continuous form of life tables, the probability of surviving to age $a$ can also be interpreted as the number of survivors at age $a$. Thus, the ratio $\delta(a)/p(a)$ indicates the relative importance of the sex difference in the number of survivors at age $a$ to the number of survivors of both sexes combined at the same age. It would be no surprise, for example, that a small $\delta(a)$ together with an even smaller $p(a)$ may yield a high ratio.

Analogously, dividing Eq. (6) by Eq. (2) given the rate of relative change of total population as a response to a rise in $\pi_m$:

$$\frac{d \ln N(t)}{d\pi_m} = -\frac{e_f^o - e_m^o}{e^o}. \tag{8}$$

meaning that, for populations with $e_f^o > e_m^o$, the relatively decline of total population equal to the ratio of sex difference in life expectancy at birth to life expectancy of the population both sexes combined.

More generally, we consider the subpopulation above a certain age

$$N_{x+}(t) = \int_x^\omega N(a,t) da$$

The relative change of $N_{x+}(t)$ with respect to $\pi_m$ is given by

$$\frac{d \ln N_{x+}(t)}{d\pi_m} = -\frac{\int_x^\omega \delta(a) da}{\int_x^\omega p(a) da} = -R_x, \tag{9}$$

where $R_x = \frac{\int_x^\omega \delta(a) da}{\int_x^\omega p(a) da}$ is the ratio of cumulative sex difference in survival above age $x$ to the cumulative survival above age $x$ for the total population. When $x=0$, Eq. (9) simplifies to Eq. (8).

It turns out that $R_x$ is a monotonically increasing function of age $x$ (See Appendix for detailed proof)

$$\frac{dR_x}{dx} > 0. \tag{10}$$

For two ages, $x_1 < x_2$, we have $R_{x_1} < R_{x_2}$, which means that, given a rise in $\pi_m$, the subpopulation above a higher age decreases at a faster rate than the subpopulation above a lower age does.

Figure 1 portrays the age trajectories of $p(a)$, $\delta(a)$, $\delta(a)/p(a)$ and $R_x$. Before middle age (around 50 years), $\delta(a)$ is rather small (solid line in red), while $p(a)$ remains at a high level (dotted line in green), so their ratio (dashed line in blue) is small. Then, $\delta(a)$ increases with age and peaks at approximately age 75 with a subsequent slow decline. At the same time, $p(a)$ declines fast at old ages with high mortality. Consequently, $\delta(x)/p(x)$ rises rapidly after middle age, meaning that the impact of a rise in $\pi_m$ on population size increases rapidly with age. $R_x$ shows a upward trend similar to $\delta(x)/p(x)$, but at a higher level, and both converge at the advanced ages.

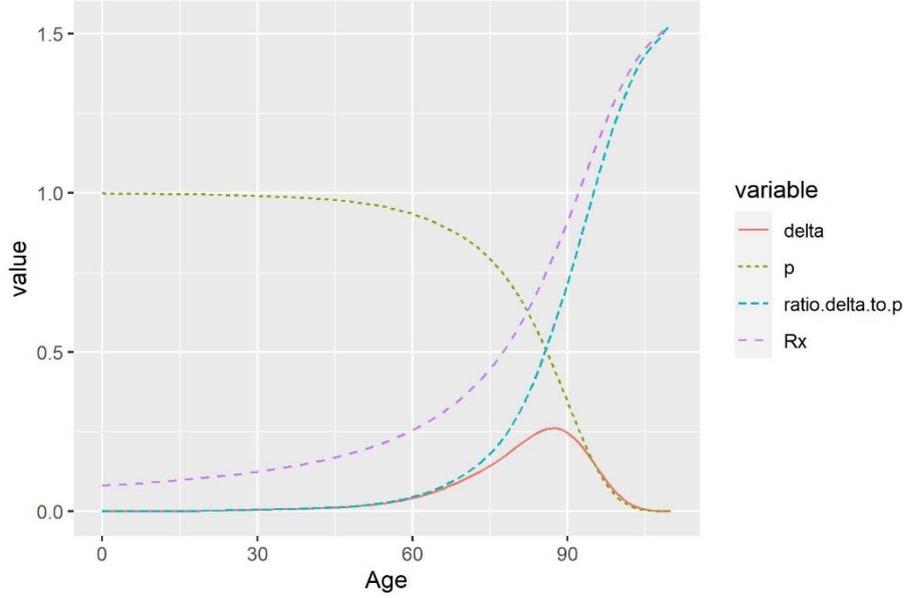

**Fig. 1** Age trajectories of relevant functions. Japan, 2010. Source: Human Mortality Database (2019).

It is assumed that a constant number of births implies that the fertility rates must rise to some degree to offset the reduction in births resulting from the rise in the SRB. As noted above, given the number of births, a higher proportion of male births implies fewer female births, and, hence, fewer women of child-bearing age in the next decades. If the fertility schedules remain unchanged, the decrease of those women will further lead to fewer births, nearly half of them being girls. Such processes will continue across generations, leading to a decreasing number of annual births, which conflicts with the assumption of an unchanging number of births. Therefore, fertility must rise to ensure a constant flow of annual births.

## 4. SRB-induced population aging

### 4.1 Case of fixed annual births

We first prove that a female-dominant stable population with a higher SRB may be younger than one with a lower SRB. We then visualize how the rise in the SRB can cause the population to age.

Another widely used indicator of population aging is the proportion of the population above age 65:

$$C_{65}(t) = \frac{\int_{65}^{\omega} N(a,t)da}{\int_{0}^{\omega} N(a,t)da} = \frac{N_{65+}(t)}{N(t)} \,. \tag{11}$$

The relative change $C_{65}(t)$ with respect to $\pi_m$ is obtained by

$$\frac{d \ln C_{65}(t)}{d\pi_m} = \frac{d \ln N_{65+}(t)}{d\pi_m} - \frac{d \ln N(t)}{d\pi_m}$$

$$= \frac{1}{N_{65+}(t)} \int_{65}^{\omega} \frac{dN(a,t)}{d\pi_m} da - \frac{1}{N(t)} \int_{0}^{\omega} \frac{dN(a,t)}{d\pi_m} da$$

$$= -\frac{\int_{65}^{\omega} \delta(a)da}{\int_{65}^{\omega} p(a)da} - \left(-\frac{\int_{0}^{\omega} \delta(a)da}{\int_{0}^{\omega} p(a)da}\right)$$

$$= -R_{65} + R_0 \tag{12}$$

Since $R_x$ is an increasing function of age $x$, $R_{65} > R_0$ and the right-hand side of Eq. (12) is negative. That is, a rise in $\pi_m$ can make the population younger. The underlying reason lies in that, as more boys are born because of masculinity of births, more of them will die before reaching old age, leading to a relatively small old age population. The total population size decreases too, but at a relatively slow rate, so the proportion of old population decreases.

Another lesson Eq. (12) offers is that the change in $C_{65}$ can be decomposed into two components, one of which is the growth of the people above age 65 and another the growth of the total population. How fast one component grows compared to another can determine how $C_{65}$ will change.

The same logic can apply to the old-age dependence ratio (OADR), another popular indicator of population aging:

$$OADR(t) = \frac{\int_{65}^{\omega} N(a,t)da}{\int_{20}^{64} N(a,t)da} = \frac{N_{65+}(t)}{{}_{45}N_{20}(t)}$$

where the denominator is the labor force of a population. Similarly, we can have

$$\frac{d \ln OADR(t)}{d\pi_m} = -\frac{\int_{65}^{\omega} \delta(a)da}{\int_{65}^{\omega} p(a)da} - \left(-\frac{\int_{20}^{64} \delta(a)da}{\int_{20}^{64} p(a)da}\right) = -R_{65} + {}_{45}R_{20}.$$

Since ${}_{45}R_{20} < R_{20} < R_{65}$, the SRB rise can decrease old age burden.

If a population with a higher SRB is younger than one with a lower SRB, how can the rise in the SRB provoke population aging? The answer is that, during the transition from a lower SRB to a higher SRB, the population composition keeps changing in terms of the cohorts born with different SRBs (Fig. 2).

Figure 2 illustrates the evolution of population composition due to the rise in the SRB. Suppose that, at time $\tau$, the proportion of male births rises from the normal $\pi_m$ to an abnormally high level $\pi_m^*$ with $\pi_m^* > \pi_m$. The resultant change in the population age structure will go through three stages until the last individual born with $\pi$ dies.

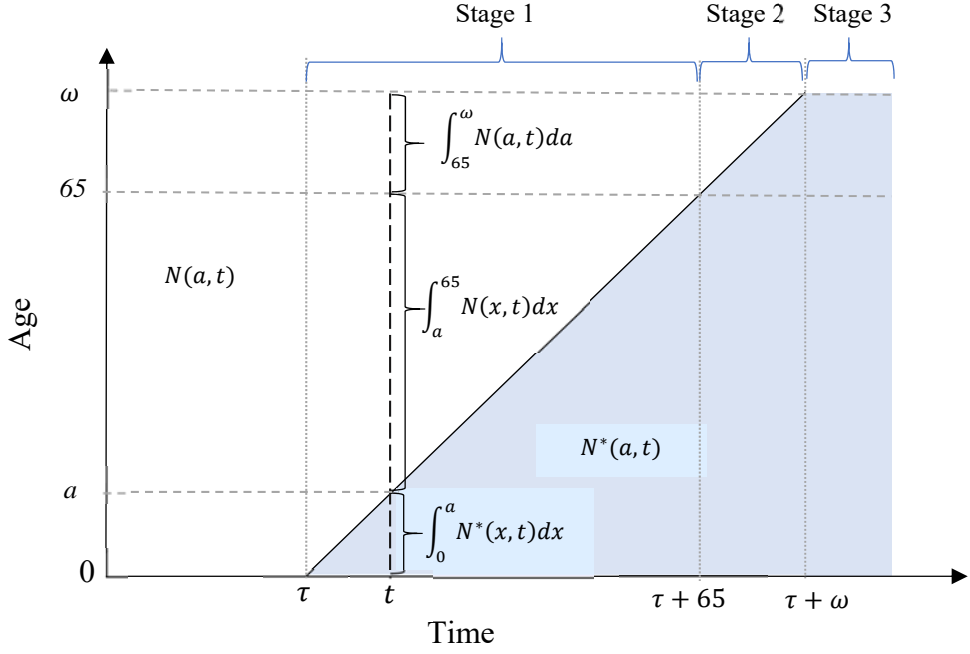

**Fig. 2** Illustration of how the rise in the SRB alters population composition and causes the population to age. Blank area on left-hand side of diagonal line consists of the cohorts born with the normal $\pi$ before time $\tau$, denoted $N(a,t)$; area shaded in light blue on right-hand side of the diagonal line represents the cohorts born with an abnormally high $\pi_m^*$ with $\pi_m^* > \pi_m$ after time $\tau$, denoted $N^*(a,t)$. Cohorts in shaded area are smaller in size than that in blank area at the same age, as shown in Eqs. (4) and (8).

In the first stage over the period from $\tau$ to $\tau + 65$, the population above age 65, born at least 65 years earlier, is not affected by $\pi_m^*$, so it has the same size as before. But the cohorts born with $\pi_m^*$ have a smaller size than those born before time $\tau$, thereby declining the total population size, $N(t)$. As such, the proportion of the elderly will rise. In this stage, the percentage of people aged 65 is given by

$$C_{65}(t) = \frac{\int_{65}^{\omega} N(a,t)da}{\int_{0}^{t-\tau} N^*(a,t)da + \int_{t-\tau}^{\omega} N(a,t)da},$$

where $N^*(a,t)$ stands for the cohorts born with the new $\pi_m^*$ and $N(a,t)$ for the cohorts with original $\pi_m$. The first term of the denominator is smaller than before time $\tau$, so $C_{65}$ becomes higher. Moreover, as time goes by, more cohorts are born with $\pi_m^*$ into the population and the first item will take an increasing share so that $C_{65}$ will continue to rise until time $t = \tau + 65$ when the population under age 65 consists all of the cohorts born with $\pi_m^*$. In this stage, the first component of Eq. (13) is zero and the equation reduces to

$$\frac{dC_{65}(t)}{C_{65}(t)d\pi_m} = R_0,$$

which is the rate at which the percentage of the population above 65 rises.

In the second stage between time $\tau + 65$ and $\tau + \omega$, the cohorts born with $\pi_m^*$ will gradually step into old age, thereby decreasing the number of older people. For instance, at time $t$, $\tau + 65 < t < \tau + \omega$), the population above age 65 is

$$N_{65+}(t) = \int_{65}^{t-\tau} N^*(a,t)da + \int_{t-\tau}^{\omega} N(a,t)da. \quad (13)$$

Substituting Eq. (13) into (12), and canceling the common term, gives the relative slope in the second stage:

$$\frac{dC_{65}(t)}{C_{65}(t)d\pi_m} = -\frac{\int_{65}^{t-\tau} \lambda(a)da}{\int_{65}^{t-\tau} p(a)da} + R_0. \quad (14)$$

As the members of the cohorts born with $\pi_m^*$ gradually live into old age, the first term on the right-hand side of Eq. (14) will increasingly offset $R_0$, leading to a slowdown of population aging. At a certain point in time, the first term will exceed $R_0$, because its absolute value approaches $R_{65}$, which is always greater than $R_0$

In the last stage, at time $\tau + \omega$, the population consist all of the cohorts born with $\pi_m^*$, and the percentage of older population will be constant again, at a level defined by Eq. (12). Since $\pi_m < \pi_m^*$, the population consists of more males than any time before. Hence, male survival disadvantages imply fewer old age survival from the births given with $\pi^*$. In other words, the population fixed number of births and higher SRB must be younger, as we shall see in the following simulation.

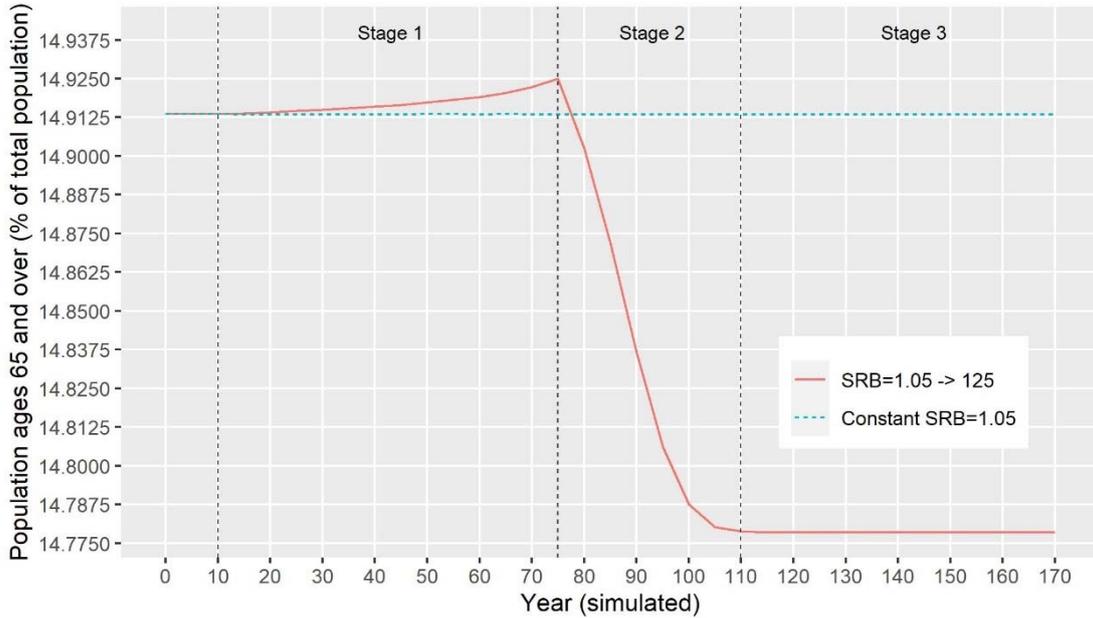

**Fig. 3** Simulated evolution of population aging with SRB perturbation. The SRB jumps from 1.05 to 1.25 at year 10 (simulated). The number of annual births is 10,000. The mortality rates for both sexes are assumed to be constant over time, with about 4 years difference in life expectancy at birth.

We illustrate the SRB-induced population aging by a female-dominant projection. The number of births is estimated by applying the fertility rate to women only (Preston et al. 2001:121-124). The data on mortality and fertility are taken from the estimates for the 1990

Zhen Zhang and Qiang LiChina by World Population Prospects (United Nations 2019). Detailed tables containing these data are available in the Appendix.

Suppose that the SRB remains at 1.05 until year 10 (simulated), followed by a rise to 1.25. The percentage of the elderly aged 65 and over (solid line in red) increases and peaks at the end of Stage 1. In Stage 2, the cohorts born with the SRB of 1.25 take an increasing share and decrease the size of the elderly population. As noted above, the reduction in population size resulting from the SRB rise is much more significant at older ages, so the proportions at old age decline rapidly. At the beginning of Stage 3, all members of the population are born with the new SRB, so the population reaches a new stationary state with a lower proportion at old age than the stationary population with the normal SRB of 1.05.

It may be noted that, despite the clear trend in population aging, the impact of the SRB on population structure does not seem as significant as expected. Indeed, the fraction above age 65 varies just in a small range (14.78%–14.93%). Regarding the apparent unimportance of the SRB, we should bear in mind that, to ensure a fixed number of births, the fertility level indeed increases accordingly. In this case, the total fertility rate (TFR) rises from 2.18 to 2.40 to have a constant flow of annual births. As we know, fertility increases can make the population younger, so it is no surprise that SRB-induced population aging is thus offset to a certain extent.

Although SRB-induced population aging may be offset by fertility, the SRB and fertility affect population structure differently. The SRB can directly change the number of girls, the future fecundity of a population, while fertility rates determine the total number of births, nearly half of whom are female, and thus indirectly impact the fecundity. When the total number of births is fixed, as in the above case, the SRB and fertility can become entangled with each other.

The different pathways of the impact of the SRB and fertility imply that they have separable influences on the growth rate, as shown in the following relationship (Coale 1972:18-22; Dublin and Lotka 1925; Preston et al. 2001:152-154):

$$r = \frac{\ln TFR + \ln \pi_f + \ln p_f(A_M)}{T}, \quad (15)$$

where $\pi_f$ gives the proportion of female births, $TFR$ is the total fertility rates, $p_f(A_M)$ the probability that women survive from birth to the mean age at childbearing $(A_M)$, and $T$ the mean length of a generation. The intrinsic growth rate is an additive function of the log of $TFR$, $\pi_f$ or $p_f(A_M)$. Hence, the effect on the growth rate depends only on the proportionate change in either of the three variables. Hence, the change in $r$ due to $\pi_f$ is obtained by

$$\Delta r = \frac{\ln(\pi_f^*/\pi_f)}{T}. \quad (16a)$$

Likewise,

$$\Delta r = \frac{\ln(TFR^*/TFR)}{T} \quad \text{or} \quad \Delta r = \frac{\ln(p_f^*(A_M)/p_f(A_M))}{T}, \quad (16b)$$

where the supper script * indicates the new level of $TFR$ and $p_f(A_M)$.

Furthermore, it follows from Eq. (16) that the three variables can separately yield the same change of the growth, though they operate differently in population dynamics. For instance, if the SRB rises from 1.05 to 1.25, or equivalently $\pi_f$ falls by 9.02% from 0.488 to 0.444 and if SRB-induced populating aging$T = 27$, then the growth rate will decline by 0.35%. The same effect can be achieved by TFR being reduced by the same 9.02% from 2.10, the replacement level of fertility, to 1.91. Analogous change of $p(A_M)$ is a decrease from 0.9844 to 0.8956, which is equivalent to Chinese women's $p(25)$ rolling back from 1990 to 1964.

Noteworthy, all the relations derived above assume that females are advantageous over males with respect to survival, i.e., $p_f(a) > p_m(a)$. In some contexts, however, such female advantage can be weakened by discriminatory behavior against girls (Guilmoto 2012a). Aa a result, female mortality rates can be higher than expected or even higher than equivalent male rates. In those cases, males have survival advantage, rather than disadvantage, i.e., $p_f(a) < p_m(a)$. Of ten countries with female child higher than expected (Alkema et al. 2014), China and India are also on the list of 12 countries with significant SRB inflation (Chao et al. 2019). In China, excess female mortality has decreased since the 1990s (Alkema et al. 2014), and, $p_f(a) > p_m(a)$ has held since then (United Nations, 2019). Indian girl child mortality has long been higher than boy child mortality, and in recent decades, female excess mortality has even worsened (Alkema et al. 2014) and will last throughout in the rest of this century (United Nation 2019). Hence, in the case of India, Eq. (4) will be positive for young age with $p_f(a) < p_m(a)$, indicating that a rise in the SRB will increase the number of male births, and further make the population younger.

*4.2 Case of a fixed regime of fertility*

The assumption of fixed births allows us to readily demonstrate how the SRB rise alters population age structure. This assumption, however, is an over-simplification. Hence, we relax the assumption on the fixed number of births by assuming a time-invariant fertility schedule.

The number of births is given by applying fertility rates to women:

$$B(t) = N(0,t) = \int_\alpha^\beta N_f(a,t) \cdot m(a) da = \int_\alpha^\beta N_f(0,t) \cdot e^{-ra} \cdot p_f(a) \cdot m(a) da, \quad (17)$$

The intrinsic growth rate $r$ is defined by female fertility and mortality schedules:

$$1 = \int_\alpha^\beta e^{-ra} \cdot p_f(a) \cdot m_f(a) da, \quad (18)$$

where $p_f(a)$ and $m_f(a)$ apply only to the females (Lotka 1939). The same growth rate can also apply to the male sub-population because the births are determined by female fertility and mortality in a female-dominant population (Coale 1972:54). Substituting $N_f(0,t) = N(0,t) \cdot \pi_f$ into Eq. (17) gives

$$N(0,t) = \int_\alpha^\beta N(0,t) \cdot \pi_f \cdot p_f(a) \cdot e^{-ra} \cdot m(a) da.$$

Canceling the $N(0,t)$ from both sides yields

$$1 = \pi_f \cdot \int_\alpha^\beta e^{-ra} \cdot p_f(a) \cdot m(a) da. \quad (19)$$

The impact of the SRB on population growth can be examined using perturbation analysis. The sensitivity of $r$ to $\pi_f$ is obtained as the derivative of $r$ with respect to $\pi_f$. If the SRB or the proportion of female births is constant with the age of women, i.e., $m_f(a) = \pi_f \cdot m(a)$, then Eq. (18), which holds only for a one-sex (female) population can be extended to Eq. (19), which holds for a two-sex population. Note that this extension can be done only for the female-dominant population because the intrinsic growth rate of the male sub-population is identical to that of the female sub-population, and thus of the total population.

We designate the integral at the right-hand side of Eq. (19) as the function $\psi(r)$. According to the rules of implicit function and chain differentiation, we have

$$\frac{d\psi}{d\pi_f} = \frac{\partial \psi}{\partial r}\frac{\partial r}{\partial \pi_f} + \frac{\partial \psi}{\partial \pi_f}. \tag{20}$$

Since $\psi$ is always equal to 1 and would not change with $\pi_f$, the left-hand side of Eq. (20) should be zero. Rearranging the right-hand side of Eq. (20) yields

$$\frac{dr}{d\pi_f} = -\frac{\frac{\partial \psi}{\partial \pi_f}}{\frac{\partial \psi}{\partial r}} = \frac{\int_\alpha^\beta e^{-ra} \cdot p_f(a) \cdot m(a) da}{\int_\alpha^\beta \pi_f \cdot a \cdot e^{-ra} \cdot p_f(a) \cdot m(a) da} = \frac{1}{\pi_f A_B} > 0, \tag{21}$$

where $A_B = \int_\alpha^\beta a \cdot e^{-ra} \cdot p_f(a) \cdot m(a) da$ is the mean age of childbearing in the stable population. In Eq. (21), $dr/d\pi_f > 0$ confirms the above findings that an increase in the proportion of female births can increase population growth and vice versa. In a female-dominant population, the more girls, the greater the fecundity – which will boost the population growth even if the TFR stays constant.

In a stable population the age distribution depends on the mortality regime and the intrinsic rate of increase:

$$c(a) = \frac{e^{-ra}p(a)}{\int_0^\omega e^{-ra}p(a)da}.$$

The proportion of the population above 65 is

$$C_{65} = \int_{65}^\omega c(a)da = \frac{\int_{65}^\omega e^{-ra}p(a)da}{\int_0^\omega e^{-ra}p(a)da}.$$

Similarly, we have

$$\frac{d\ln C_{65}}{dr} = A_p - A_{65}, \tag{22}$$

where $A_p = \int_0^\omega a \cdot c(a)da / \int_0^\omega c(a)da$ is the mean age of the stable population, and $A_{65} = \int_{65}^\omega a \cdot c(a)da / \int_{65}^\omega c(a)da$ is the mean age of the population above age 65. Since $A_p$ is always lower than $A_{65}$, Eq. (22) is negative, indicating that a growing population tends to be young.

Multiplying Eq. (22) by (21) yields

$$\frac{d\ln C_{65}}{d\pi_f} = \frac{A_p - A_{65}}{\pi_f A_B}. \tag{23}$$

Further, by multiplying both sides of Eq. (23) by $\pi_f$ and rearranging gives us the elasticity of $C_{65}$ with respect to $\pi_f$:

$$\frac{d \ln C_{65}}{d \ln \pi_f} = \frac{A_p - A_{65}}{A_B}. \tag{24}$$

Since $A_p < A_{65}$, Eq. (24) is negative, indicating how much $C_{65}$ would fall due to a proportional increase in $\pi_f$.

One important difference between Eqs. (12) and (23) or (24) should be noted that, with the assumption of fixed births, sex differences in survival can affect the results, but in the case of fixed fertility, sex difference in survival is not involved, as shown in Eq. (23), so the impact of the SRB on population age structure will not change even with $p_f(a) < p_m(a)$. The reason behind such difference is that, given the number of births, not only the SRB but also sex difference in survival determine how many extra male births will be born and can survive to certain ages, while more male births come at compensate of female births. But given the fertility level, in a female-dominant population, the number of births only depends on women of childbearing and fertility rates. Hence, in the case of fixed fertility, even a reversal of sex difference in survival will not change the trend in population aging that is determined by female vital rates.

Figure 4 illustrates the evolution of population aging given a perturbation of the SRB. Fertility rates are adjusted to annual births number of 10,000 based on the age pattern of fertility rates in 1990 China so that the results can be comparable to the above case of a fixed number of births (Fig. 3). The SRB remains at the normal level of 1.05 in the first 10 years (simulated). Likewise, suppose that from year 10 and onward, the SRB rises from 1.05 to 1.25. As expected, the proportion at old age rises rapidly and peaks at approximately 16.8%, followed by decreasing frequency oscillations while approaching a new level. The oscillations arise from the perturbation in the number of male and female births due to the SRB rise (Fig. 5), as discussed in Coale (1972:63-65). During the convergence to a stable form, the slowly damped oscillations in the percentage at old age remain still visible until around year 200, indicating the long-run impact of a rise in the SRB.

It follows from Eq. (16) that different, but equivalent, changes in the SRB, fertility and mortality can separately lead to the same change of the growth rate. It thus might be interesting to ask whether those equivalent changes have the same effect on population age distribution, and if now, how different they are? Since there are no analytic expressions for the impact on $C_{65}$ of changes in fertility and mortality, as in Eq. (23) for the SRB rise, we use simulations to investigate the questions. Consider two scenarios, first of which is that $p(A_M)$ is assumed to fall by 9.02% at year 10 while the SRB remains 1.05 and the TFR stays constant. The second is that the TFR drops by 9.02% at year 10 while the values of the SRB and $p(A_M)$ remain unchanging.

As shown in Fig. 4, the decline of TFR brings about the largest rise in $C_{65}$, slightly higher than the SRB-induce rise in $C_{65}$, while the effect of the reduction of $p(A_M)$ is the smallest. Besides, the three factors differ in the time to attain a new equilibrium after a perturbation in them: about 35 years for $p(A_M)$, 70 years for the TFR and nearly 100 years for SRB.

The simulation results echo previous findings that fertility decline is most important for making population old (e.g., Lee and Zhou 2017). Furthermore, we find that the SRB rise is also an important factor of population aging, compared to either TFR or $p(A_M)$. Nevertheless, the effect of either of TFR, SRB or $p(A_M)$ is not so marked, indeed. For instance, the TFR dropping from 2.10 to 1.91 only results in a rise of two percentages in $C_{65}$. This seems contrary to our impression that world populations, mostly from developing countries, have been aging rapidly. One reason behind is that, in actual fact, population aging is driven by a combined effects of fertility decline, mortality improvement, and in some countries, birth masculinity. Though those factors have separate influences on the population aging, they can reinforce one another through population dynamics. It turns out that the combination of changes in the three factors can raise $C_{65}$ by 4.89 percentages, more than 4.60 percentages, the sum of respective increase in $C_{65}$ resulting from each of the three factors (Fig. A1).

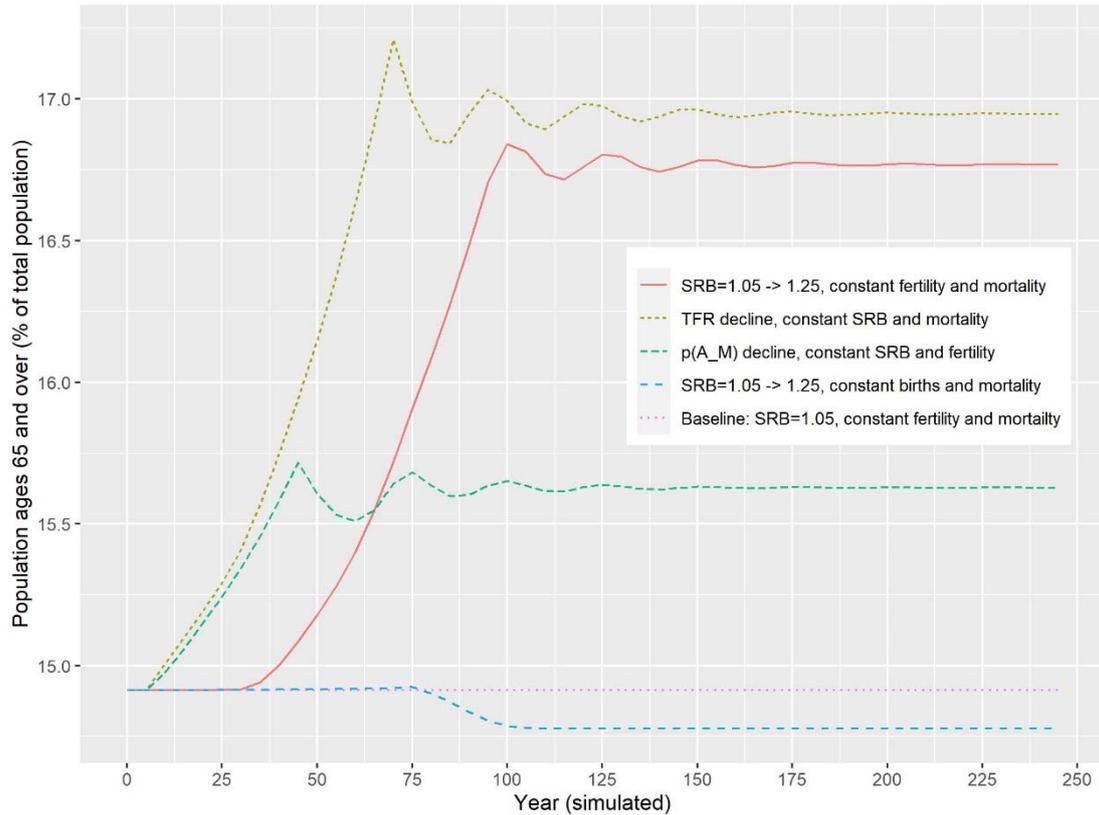

**Fig. 4** Trajectories of percentage of elderly aged 65 and over. Setting for mortality rates are the same as in Fig. 3.

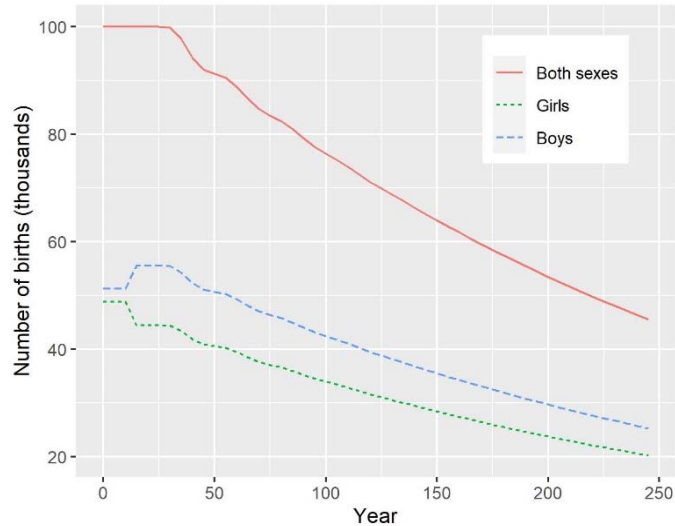

**Fig. 5** Simulated numbers of births for females, males, and both sexes combined.

## 5. Conclusions

This study investigates the role of the SRB in population dynamics, emphasizing the change of age distribution. With a simple framework derived from classical stable population models, we reveal that the change in the SRB can affect the size, structure, and growth of a population. Typically, the rise in the SRB can make population old and vice versa.

The underlying mechanism is intuitively straightforward. A rise in the SRB will introduce into the population more boys subject to higher mortality, implying more deaths of the new, young cohorts. Moreover, a higher proportion of male births implies fewer girls to be born, and, hence, fewer women that will give birth in the future. Since the decrease in the number of childbearing women will further reduce the number of births in the next generation, such a process will continue across generations until the population eventually tends to be stable more than one century later. If the number of births is assumed to be fixed, fertility needs to rise to some degree to ensure a constant flow of births. If fertility is assumed to be unchanging, the number of births will decline and young generations will become smaller. In either case, the transient period from a stationary or stable population with a normal SRB to the next one with an abnormally high SRB, the population consists of two parts: old cohorts born with a normal SRB and young cohorts born with the new SRB. Since the new, young cohorts are smaller in size than what it would be if with a normal SRB, the population will be older than before the SRB rises. After the new equilibrium is attained, the population age structure can be younger or older than before the rise in the SRB, depending on whether a fixed number of births (implicitly changing fertility) or fixed fertility regime is assumed.

It should be noted that this study has several limitations. First, our analysis is based on a two-sex model with the assumption of female dominance. Most, if not all, countries with abnormally high SRB have excess males, so the assumption of female dominance generally holds. However, since marriage is usually a precondition of childbearing in some contexts, marriage squeeze may affect the extent to which the assumption of female dominance is satisfied. Hence, a more realistic modelling could consider marriage pattern (e.g., Guilmoto 2012b). Second, in this study, we mainly consider two scenarios on fertility, fixed number of births and fixed fertility regimes. In fact, the SRB inflation is often accompanied with fertility decline, particularly at the early stage of demographic transition. Hence, our current framework could be extended to embrace fertility decline.

## 6. Acknowledgments

This study was funded by the National Natural Science Foundation of China (71473044) and Shanghai Planning Office of Philosophy and Social Science(2019BSH002).

# Appendices

**Appendix 1. Proof of Eq. (10).**

Consider $R_x = \frac{\int_x^\omega \delta(a)da}{\int_x^\omega p(a)da}$, then $\frac{dR_x}{dx} > 0$.

*Proof:*

Differentiating $R_x$ with respect to $\pi_m$ is given by

$$\frac{dR_x}{dx} = \frac{1}{\left(\int_x^\omega p(a)da\right)^2}\left(-\delta(x)\int_x^\omega p(a)da + p(x)\int_x^\omega \delta(a)da\right)$$

Substituting $\delta(x) = p_f(x) - p_m(x)$ and $p(x) = \pi_m p_m(x) + (1-\pi_m)p_f(x)$ into the right-hand side of the equation, and algebra arranging yields the right-hand side of the equation is

$$\frac{dR_x}{dx} = \frac{1}{\left(\int_x^\omega p(a)da\right)^2} p_f(x)p_m(x)\left(e_f^o(x) - e_m^o(x)\right).$$

Since $e_f^o(x) > e_m^o(x)$, we have $\frac{dR_x}{dx} > 0$, meaning that $R_x$ is a monotonically increasing function of ag $x$.

**Appendix 2. Table S1 Life tables for males and females**

| | Male | | | | | | Female | | | | |
|---|---|---|---|---|---|---|---|---|---|---|---|
| x | nqx | lx | nLx | Tx | ex | x | nqx | lx | nLx | Tx | ex |
| 0 | 0.0442 | 1.0000 | 0.9624 | 67.43 | 67.43 | 0 | 0.0381 | 1.0000 | 0.9671 | 71.43 | 71.43 |
| 1 | 0.0107 | 0.9558 | 3.7977 | 66.47 | 69.54 | 1 | 0.0095 | 0.9619 | 3.8246 | 70.46 | 73.25 |
| 5 | 0.0050 | 0.9456 | 4.7161 | 62.67 | 66.28 | 5 | 0.0035 | 0.9527 | 4.7553 | 66.64 | 69.94 |
| 10 | 0.0030 | 0.9409 | 4.6972 | 57.96 | 61.60 | 10 | 0.0020 | 0.9494 | 4.7422 | 61.88 | 65.18 |
| 15 | 0.0040 | 0.9380 | 4.6808 | 53.26 | 56.78 | 15 | 0.0035 | 0.9475 | 4.7292 | 57.14 | 60.31 |
| 20 | 0.0050 | 0.9343 | 4.6598 | 48.58 | 52.00 | 20 | 0.0040 | 0.9442 | 4.7115 | 52.41 | 55.51 |
| 25 | 0.0065 | 0.9296 | 4.6331 | 43.92 | 47.24 | 25 | 0.0050 | 0.9404 | 4.6903 | 47.70 | 50.72 |
| 30 | 0.0075 | 0.9236 | 4.6008 | 39.29 | 42.54 | 30 | 0.0060 | 0.9357 | 4.6646 | 43.01 | 45.96 |
| 35 | 0.0100 | 0.9167 | 4.5607 | 34.69 | 37.84 | 35 | 0.0075 | 0.9301 | 4.6332 | 38.34 | 41.22 |
| 40 | 0.0144 | 0.9076 | 4.5053 | 30.13 | 33.19 | 40 | 0.0100 | 0.9232 | 4.5929 | 33.71 | 36.52 |
| 45 | 0.0198 | 0.8945 | 4.4283 | 25.62 | 28.64 | 45 | 0.0154 | 0.9140 | 4.5348 | 29.12 | 31.86 |
| 50 | 0.0363 | 0.8768 | 4.3044 | 21.19 | 24.17 | 50 | 0.0242 | 0.8999 | 4.4452 | 24.58 | 27.32 |
| 55 | 0.0583 | 0.8450 | 4.1017 | 16.89 | 19.99 | 55 | 0.0383 | 0.8781 | 4.3068 | 20.14 | 22.93 |
| 60 | 0.1002 | 0.7957 | 3.7793 | 12.79 | 16.07 | 60 | 0.0695 | 0.8446 | 4.0760 | 15.83 | 18.75 |
| 65 | 0.1651 | 0.7160 | 3.2844 | 9.01 | 12.58 | 65 | 0.1114 | 0.7859 | 3.7104 | 11.76 | 14.96 |
| 70 | 0.2747 | 0.5978 | 2.5782 | 5.72 | 9.57 | 70 | 0.1958 | 0.6983 | 3.1497 | 8.04 | 11.52 |
| 75 | 0.4013 | 0.4335 | 1.7327 | 3.14 | 7.25 | 75 | 0.2964 | 0.5616 | 2.3918 | 4.90 | 8.72 |
| 80 | 0.5476 | 0.2596 | 0.9425 | 1.41 | 5.44 | 80 | 0.4612 | 0.3951 | 1.5200 | 2.50 | 6.34 |
| 85 | 0.7326 | 0.1174 | 0.3721 | 0.47 | 3.99 | 85 | 0.6468 | 0.2129 | 0.7202 | 0.98 | 4.62 |
| 90 | 0.8969 | 0.0314 | 0.0866 | 0.10 | 3.08 | 90 | 0.8132 | 0.0752 | 0.2231 | 0.26 | 3.50 |

| | | | | | | | | | | |
|---|---|---|---|---|---|---|---|---|---|---|
| 95 | 1.0000 | 0.0032 | 0.0101 | 0.01 | 3.12 | 95 | 1.0000 | 0.0140 | 0.0401 | 0.04 | 2.86 |

**Appendix 3. Table S2 Age-specific fertility rates**

| Age | mx |
|---|---|
| 15-19 | 0.094 |
| 20-24 | 0.945 |
| 25-29 | 0.791 |
| 30-34 | 0.268 |
| 35-39 | 0.057 |
| 40-44 | 0.023 |
| 45-49 | 0.006 |

**Appendix 4.**

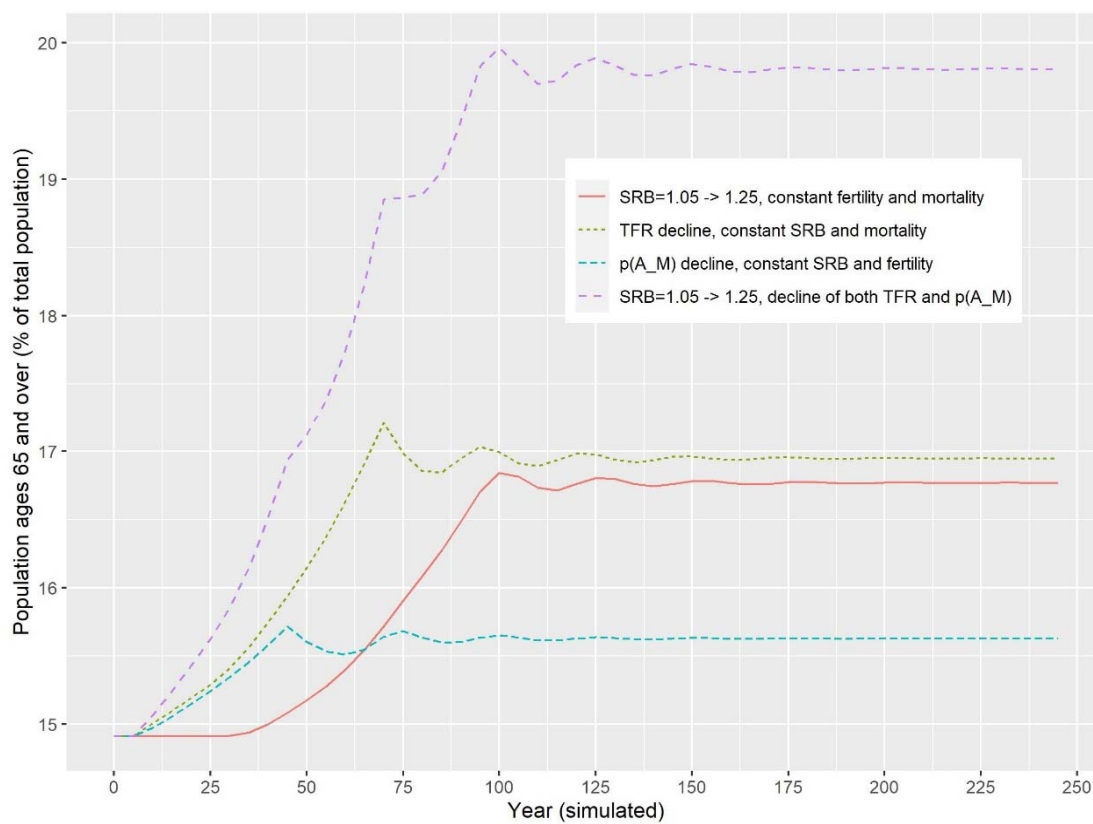

**Fig. A1** Trajectories of percentage of elderly aged 65 and over, given the perturbations in TFR, SRB and $p(A_M)$. For simulation assumptions, see main text. Setting for mortality rates are the same as in Fig. 3.